\documentclass{article}
\usepackage{arxiv}
\usepackage{braket}
\usepackage{listings}
\usepackage{amsmath}
\usepackage{tikz}
\usetikzlibrary{quantikz2}

\title{Bridging Quantum Mechanics and Computing: A Primer for Software Engineers}

\author{
 Arvind W Kiwelekar \\
  Department of Computer Engineering\\
  Dr. Babasaheb Ambedkar Technological University \\
  Lonere-402103 Dist Raigad India \\
  \texttt{awk@dbatu.ac.in} \\ 
}

\begin{document}

\maketitle
\begin{abstract}
Quantum mechanics, the fundamental theory that governs the behavior of matter and energy at microscopic scales, forms the foundation of quantum computing and quantum information science. As quantum technologies progress, software engineers must develop a conceptual understanding of quantum mechanics to grasp its implications for computing. This article provides a primer on fundamental quantum mechanics principles pertinent to software engineers, including wave-particle duality, superposition, entanglement, quantum states, and quantum measurement. Unlike traditional physics-oriented discussions, this article focuses on computational perspectives, assisting software professionals in bridging the gap between classical computing and emerging quantum paradigms. 
\end{abstract}

\keywords{Quantum Mechanics \and  Quantum Computing \and
 Quantum Measurement and Superposition \and 
Quantum Entanglement and Nonlocality\and  Classical vs. Quantum Computing }

\tableofcontents
\section{Introduction}
Quantum Computing (QC) is a specific application of Quantum Mechanics (QM) that focuses on developing computers utilizing quantum bits a.k.a {\em qubits} instead of classical bits. Qubits can exist in superpositions of states, allowing quantum computers to perform multiple calculations simultaneously, leading to significant speedups for specific computational problems. For instance, Shor's algorithm for factoring large numbers and Grover's algorithm for searching unsorted databases are notable quantum algorithms that exploit these quantum properties\cite{10.1007/s00453-002-0971-8}. The advent of quantum computing represents a paradigm shift in computational capabilities, as it employs the principles of quantum mechanics to solve problems intractable for classical computers\cite{10.1007/s00453-002-0971-8}. 

The quantum revolution enables the design of novel techniques for building information systems referred to as Quantum Information Systems(QIS), including more secure communication models \cite{10.1007/s00453-002-0971-8}, quantum cryptography and quantum teleportation\cite{10.1038/nature07127}.   QIS, which includes quantum computing and quantum communication, focuses on the processing, transmitting, and storing information using quantum states. By integrating quantum mechanics and information science,  QIS aims to generalize classical information systems, leading to advances in designing novel applications \cite{10.1038/nature07127}.

The foundational theories for quantum computing and information systems are deeply rooted in the principles of quantum mechanics. Table \ref{t1} describes these fields' main concepts and focus. Concepts such as superposition, entanglement, and contextuality are essential for understanding how quantum systems can be harnessed for computational purposes. A solid grasp of these foundational theories is crucial for developing practical quantum algorithms, designing robust quantum systems, and exploring the vast potential of quantum computing to solve complex problems.

Software engineers are well-versed in classical algorithms and data structures but often lack the foundational understanding of quantum mechanics needed to grasp quantum algorithms. This paper bridges that gap by introducing key quantum concepts relevant to computing.
A major challenge in adopting quantum technologies for software engineering lies in understanding the interconnections between Quantum Mechanics (QM), Quantum Computing (QC), and Quantum Information Systems (QIS). For instance, the quantum measurement problem complicates the interpretation of quantum states post-measurement. Existing literature often assumes prior knowledge of quantum principles, making it difficult for newcomers to build a solid foundation. By systematically outlining essential quantum phenomena, this paper facilitates a smoother transition for software engineers into the world of quantum computing.

\begin{table}[t]
    \centering
    \begin{tabular}{|p{1.25in}|p{1.25in}|p{1.25in}|p{1.25in}|}
\hline 
     \textbf{Field}	& \textbf{Focus}	& \textbf{Main Concepts}	&\textbf{Example Applications} \\ \hline 
Quantum Mechanics (QM)	& Fundamental physics &	Wavefunction, superposition, entanglement	& Atomic structure, quantum chemistry \\ \hline 
Quantum Information Systems (QIS) &	Quantum-based information processing	&Qubits, quantum communication, entanglement  &	Quantum cryptography, quantum networks \\ \hline 
Quantum Computing (QC) &	Computation using quantum mechanics&	Quantum gates, quantum circuits, algorithms	& Factoring large numbers (Shor’s algorithm), AI applications \\ \hline 

    \end{tabular}
    \caption{Comparing Quantum Mechanics, Quantum Computing and Quantum Information Systems}
    \label{t1}
\end{table}

\section{Quantum Computing: Overview}
Quantum computing represents a paradigm shift in computational technology, leveraging the principles of quantum mechanics to process information in fundamentally different ways compared to classical computing. This section provides an overview of key concepts in quantum computing, including qubits, quantum gates, quantum circuits, entanglement, measurement, parallelism, and interference.

A {\em qubit}, or quantum bit, is the fundamental unit of quantum information. Unlike a classical bit, which can be either 0 or 1, a qubit can exist in a superposition of both states simultaneously. Mathematically, a qubit can be represented as:

\[
\ket{\psi} = \alpha |0\rangle + \beta |1\rangle
\]

where \( \alpha \) and \( \beta \) are complex coefficients satisfying the normalization condition \( |\alpha|^2 + |\beta|^2 = 1 \).

The ability of qubits to exist in superposition allows quantum computers to perform multiple calculations simultaneously, leading to potential speedups for certain problems, such as factoring large integers and searching unsorted databases \cite{10.48550/arxiv.1712.09289}.  
\begin{table}[!ht]
    \centering
    \begin{tabular}{|p{1.6in}|p{1.75in}|p{1.25in}|}
    \hline
        \textbf{Symbol} &  \textbf{Application} &  \textbf{Matrix}\\ \hline
           
        \begin{quantikz}
        &\gate{X}&
    \end{quantikz}  & 
    $X\ket{0} = \ket{1}$ \newline
    $X\ket{1} = \ket{0}$ \newline
    & 
    \[
    \begin{pmatrix}
        0 & 1 \\
        1  & 0 \\ 
    \end{pmatrix}\]
    \\  \hline

    \begin{quantikz}
        &\gate{Y}&
    \end{quantikz} & 
    $Y\ket{0} = i\ket{1}$ \newline
    $Y\ket{1} = -i\ket{0}$ \newline
    & 
    \[\begin{pmatrix}
        0 & -i \\
        i  & 0 \\ 
    \end{pmatrix}\]
    \\  \hline

    \begin{quantikz}
        &\gate{Z}&
    \end{quantikz} & 
    $Z\ket{0} = \ket{0}$ \newline
    $Z\ket{1} = -\ket{1}$ \newline
    & 
    \[\begin{pmatrix}
        1 & 0 \\
        0  & -1\\ 
    \end{pmatrix}\]
    \\  \hline

 \begin{quantikz}
        &\gate{S}&
    \end{quantikz} & 
    $S\ket{0} = \ket{1}$ \newline
    $S\ket{1} = i\ket{0}$ \newline
    & 
    \[\begin{pmatrix}
        1 & 0 \\
        0  & i \\ 
    \end{pmatrix}\]
    \\  \hline 

   \begin{quantikz}
        &\gate{T}&
    \end{quantikz} &
    $T\ket{0} = \ket{0}$ \newline
    $T\ket{1} = e^{i\pi/4}\ket{1}$ \newline
    & \[
    \begin{pmatrix}
        1 & -0  \\
        0  &   e^{i\pi/4} \\ 
    \end{pmatrix}\]
    \\  \hline
 \begin{quantikz}
        &\gate{H}&
    \end{quantikz} & 
     $H\ket{0} =  \frac{\ket{0} + \ket{1}}{\sqrt{2}}$ \newline
    $H\ket{1} = \frac{\ket{0} - \ket{1}}{\sqrt{2}}$ \newline
    & 
     \[\frac{1}{\sqrt{2}}
     \begin{pmatrix}
        1 & 1 \\
        1  & -1\\ 
    \end{pmatrix}\]
    \\  \hline
\begin{center}
       \begin{quantikz}
       \ket{x} &\permute{2,1}&  \ket{y} \\
      \ket{y} & &  \ket{x} \\ 
    \end{quantikz}
     \end{center}
&
\begin{enumerate}
           \item $swap\ket{00} = \ket{00}$
           \item $swap\ket{01} = \ket{10}$
           \item $swap\ket{10} = \ket{01}$
           \item $swap\ket{11} = \ket{11}$     
          \end{enumerate} 

          &
 \begin{center}
        \[\begin{pmatrix}
            1 & 0 & 0 & 0  \\
            0 & 0 & 1 & 0  \\
            0 & 1 & 0 & 0  \\
            0 & 0 & 0 & 1  \\
            
        \end{pmatrix}\]
    \end{center}
 \\ \hline 
 \[
\begin{quantikz}
     \lstick{$\ket{x}$} & \ctrl{1} &    \\
     \lstick{$\ket{y}$} &  \targ{} & \\
\end{quantikz}\]

&
\begin{enumerate}
           \item $CNOT\ket{00} = \ket{00}$
           \item $CNOT\ket{01} = \ket{01}$
           \item $CNOT\ket{10} = \ket{11}$
           \item $CNOT\ket{11} = \ket{10}$     
          \end{enumerate}
          &
          \begin{center}
       \[ \begin{pmatrix}
            1 & 0 & 0 & 0  \\
            0 & 1 & 0 & 0  \\
            0 & 0 & 0 & 1  \\
            0 & 0 & 1 & 0  \\
            
        \end{pmatrix}\]
    \end{center}
\\ \hline 
    \end{tabular}
    \caption{Working of Some of the Frequently Used Quantum Gates}
    \label{gates}
\end{table}
\textit{ Quantum gates} are the basic building blocks of quantum circuits, analogous to classical logic gates. They perform operations on qubits and are represented by unitary operators these are matrices with specific properties. Common quantum gates include the Hadamard gate (H), Pauli-X gate (NOT), and CNOT gate (Controlled-NOT). Table \ref{gates} describes the working of some of the frequently used quantum gates along with their symbols and corresponding matrix representation. 

Quantum gates manipulate the state of qubits, enabling the implementation of quantum algorithms. The combination of multiple quantum gates forms {\em quantum circuits}, which can execute complex computations. Understanding quantum gates is essential for software engineers as they design and implement quantum algorithms \cite{10.1088/0253-6102/56/1/14}.

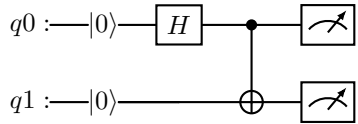
\begin{figure}
   
    \begin{center}
    \begin{quantikz}
        q0: & \ket{0} & \gate{H} & \ctrl{1} & \meter{} \\
    q1: &\ket{0} & \qw      & \targ{}  & \meter{}
   \end{quantikz} 
\end{center}
    \caption{A  quantum circuit demonstrating  gates, superposition and entaglment}
    \label{qc}
\end{figure}

A {\em quantum circuit} is a model for quantum computation that consists of a sequence of quantum gates applied to a set of qubits. Quantum circuits are typically represented using circuit diagrams, where qubits are depicted as horizontal lines and gates as boxes connecting these lines. Quantum circuits provide a visual and mathematical framework for designing quantum algorithms. They allow for the systematic representation of quantum operations and facilitate the analysis of quantum computations. The ability to construct and optimize quantum circuits is crucial for developing efficient quantum algorithms and applications \cite{10.1038/nature09418}. Figure \ref{qc} depicts a simple example of quantum circuit demonstrating  quantum gates, superposition and entanglement.  The QSkit code in Figure \ref{code} implements and simulates the same circuit.

\begin{figure}[!ht]
    \centering

         \begin{tabular}{|p{6.25in}|}
         \hline 
         \begin{small}

             \begin{lstlisting}[language=python] 
from qiskit import QuantumCircuit, Aer, transpile, assemble, execute
from qiskit.visualization import plot_histogram

# Step 1: Create a quantum circuit with
#2 qubits and 2 classical bits
qc = QuantumCircuit(2, 2)

# Step 2: Apply Hadamard gate on the first qubit 
#(creates superposition)
qc.h(0)

# Step 3: Apply CNOT gate with qubit 0 as control and 
#qubit 1 as target (creates entanglement)
qc.cx(0, 1)

# Step 4: Measure the qubits
qc.measure([0,1], [0,1])

# Step 5: Draw the circuit
print(qc.draw())

# Step 6: Simulate the circuit and get results
simulator = Aer.get_backend('aer_simulator')
tqc = transpile(qc, simulator)
qobj = assemble(tqc)
result = simulator.run(qobj).result()

# Step 7: Get and plot measurement results
counts = result.get_counts()
plot_histogram(counts)


\end{lstlisting}
\end{small} \\ \hline
       \end{tabular}

    \caption{A Sample Program in Qskit}
    \label{code}
\end{figure}

{\em Entanglement measurement} refers to assessing the degree of entanglement between quantum particles. Entangled particles exhibit correlations that classical physics cannot explain, and measuring one particle instantaneously affects the state of the other, regardless of the distance between them. Entanglement is a key resource in quantum computing and quantum information systems. It enables quantum teleportation, superdense coding, and secure quantum communication protocols. Understanding how to measure and manipulate entangled states is essential for software engineers working on quantum technologies \cite{10.47852/bonviewjdsis3202656}.

{\em Quantum parallelism} refers to the ability of quantum computers to evaluate multiple inputs simultaneously due to the superposition of qubit states. This property allows quantum algorithms to explore a vast solution space in parallel.  Quantum parallelism is a significant advantage of quantum computing, enabling algorithms like Grover's search algorithm to achieve quadratic speedup over classical counterparts. This capability has profound implications for solving complex computational problems more efficiently \cite{10.1364/ol.43.005765}.

{\em Quantum interference} occurs when the probability amplitudes of quantum states combine, leading to constructive or destructive interference effects. This phenomenon is a direct consequence of the wave-like nature of quantum states.  Interference is crucial for the success of many quantum algorithms, as it allows for the amplification of correct solutions and cancellation of incorrect ones. Understanding how to harness quantum interference is essential for optimizing quantum algorithms and enhancing their performance \cite{10.1103/physreva.85.032326}.  

\section{Classical vs Quantum Computing} Computing has long been driven by classical principles, where information is processed using binary states (0s and 1s). However, with the emergence of quantum computing, a new computational paradigm has emerged, leveraging the principles of quantum mechanics to perform tasks that are infeasible for classical systems. This section outlines the key differences between classical and quantum computing, as depicted in Table \ref{qcvscc}, and explains their implications for software engineers.

\subsection{Fundamental Unit of Information}
The fundamental unit of information is a crucial concept in both classical and quantum computing, as it defines how data is represented and manipulated within each computational paradigm. 

In classical computing, the fundamental unit of information is the \textit{
bit}. A bit can exist in one of two states: 0 or 1. These binary states are the building blocks of all classical information processing, allowing for the representation of data, execution of algorithms, and storage of information.
In quantum computing, the fundamental unit of information is the \textit{quantum bit} or qubit. Unlike classical bits, qubits can exist in a superposition of states.  The ability of qubits to exist in superposition allows quantum computers to perform multiple calculations simultaneously, leading to potential exponential speedups for certain problems. This property is a key advantage of quantum computing over classical computing.  
 \begin{table}[t]
     \centering
     \begin{tabular}{|p{1.5in}|p{2in}|p{2.5in}|}
         \hline

\textbf{Feature}	& \textbf{Classical Computing}& 	\textbf{Quantum Computing} \\ \hline 

\multicolumn{3}{|p{4in}|}{\centering{\textbf{Fundamental Unit of Information}}}    \\ \hline 
Basic Unit	& Bit (0 or 1)	& Qubit ( superposition of 0 and 1 )  \\ \hline
Representation &	Discrete states (0 or 1)&	Quantum states using wavefunctions \\ \hline
Processing Power &	Processes one state at a time &	Can exist in multiple states simultaneously \\ \hline
\multicolumn{3}{|p{4in}|}{\centering{\textbf{Processing Model}}}    \\ \hline 
Computation Type	& Deterministic (Exact outputs for the same input)	& Probabilistic (Output is a probability distribution) \\ \hline
Logic Gates	& AND, OR, NOT, XOR	& Hadamard, CNOT, Toffoli, Phase Gates \\ \hline
State Evolution	& Follows classical Boolean logic & 	Evolves based on quantum mechanics principles \\ \hline
\multicolumn{3}{|p{4in}|}{\centering{\textbf{Parallelism}} }   \\ \hline 
Parallel Processing	& Limited (multi-core/multi-threading)	& Superposition enables massive parallelism \\ \hline
State Correlation	& Independent bits	& Qubits can be entangled, meaning the state of one qubit depends on another \\ \hline
Speedup Potential &	Polynomial speedup with better hardware & 	Exponential speedup for certain problems \\ \hline

\multicolumn{3}{|p{4in}|}{\centering{\textbf{Problem Solving}}}    \\ \hline
Factorization &	Exponential complexity (RSA encryption relies on this)	 & Shor’s Algorithm solves it exponentially faster \\ \hline
Unstructured Search	 & Linear search (O(N))	& Grover’s Algorithm achieves $O(\sqrt{N})$ speedup \\ \hline
Optimization	& Iterative heuristics (gradient descent)	& Quantum Annealing offers potential quantum speedup \\ \hline
\multicolumn{3}{|p{4in}|}{\centering{\textbf{Error Tolerance}} }   \\ \hline
Error Handling &	Redundancy, error correction codes &	Quantum error correction is complex but essential \\ \hline 
Decoherence	& Not an issue &	Qubits are highly sensitive to external noise \\ \hline 
Scalability	& Scaling follows Moore’s Law &	Requires new approaches to scaling (e.g., fault-tolerant quantum computing) \\ \hline 
     \end{tabular}
     \caption{Classical vs Quantum Computing}
     \label{qcvscc}
 \end{table}
\subsection{Processing Model} The processing models of classical computing and quantum computing differ fundamentally in their approaches to computation. Classical computing relies on deterministic processes, while quantum computing embraces probabilistic computation. 

In classical computing, the processing model is deterministic, meaning that the outcome of a computation is entirely predictable based on the input values and the algorithm used. For example, a classical algorithm will produce the same output every time it is executed with the same input.
The deterministic nature of classical computing allows software engineers to design algorithms with guaranteed outcomes, making it easier to reason about program behavior and performance. However, this predictability can also limit the ability to explore multiple solutions simultaneously, particularly for problems that benefit from parallel processing.

Quantum computing operates on a probabilistic model, where the outcomes of computations are inherently uncertain and described by probability amplitudes. Quantum algorithms leverage the principles of superposition and entanglement, allowing for the simultaneous exploration of multiple computational paths. The probabilistic nature of quantum computing introduces a new paradigm for algorithm design. Software engineers must adapt to thinking in terms of probabilities and uncertainties, which can be challenging for those accustomed to deterministic models. 
\subsection{Parallelism} The concept of parallelism is a critical differentiator between classical computing and quantum computing. In classical computing, parallelism is limited by the sequential execution of operations, while quantum computing leverages the principles of superposition and entanglement to achieve a fundamentally different form of parallelism. 

Classical computing primarily relies on a deterministic model where operations are executed sequentially. Each operation is performed one after the other, which can lead to inefficiencies when dealing with complex problems that require extensive computational resources. For instance, classical algorithms often require multiple iterations to explore possible solutions, which can be time-consuming and resource-intensive.

Quantum computing fundamentally changes the notion of parallelism through the use of qubits, which can exist in superpositions of states. A qubit can represent both 0 and 1 simultaneously, allowing quantum computers to evaluate multiple possibilities at once. This property enables quantum algorithms to explore a vast solution space in parallel, significantly enhancing computational efficiency for certain problems.

In addition to superposition, quantum computing utilizes entanglement, where the states of multiple qubits become correlated. This nonlocal correlation allows quantum computers to perform operations on entangled qubits in a way that is not possible with classical bits. The ability to manipulate entangled states enables complex quantum operations that can lead to exponential speedups in computation.

The combination of superposition and entanglement gives rise to quantum parallelism, where quantum algorithms can process a large number of inputs simultaneously. For example, Grover's search algorithm can search an unsorted database in \(O(\sqrt{N})\) time, compared to the \(O(N)\) time required by classical algorithms. 

For software engineers transitioning to quantum computing, grasping the concept of quantum parallelism is essential. It requires a shift in thinking from deterministic, sequential processes to probabilistic, parallel operations.

\subsection{Problem Solving Capabilities}The problem-solving capabilities of classical computing and quantum computing differ significantly due to their underlying principles and computational models. This section compares classical and quantum algorithms in terms of their effectiveness in solving specific problems, such as factorization, unstructured search, and optimization.

\subsubsection*{Factorization} In classical computing, the most well-known algorithm for integer factorization is the general number field sieve (GNFS), which operates in sub-exponential time. However, its time complexity grows rapidly with the size of the number being factored, making it inefficient for large integers. For example, factoring a 2048-bit integer can take an impractically long time using classical methods.

Quantum computing offers a significant advantage in factorization through Shor's algorithm, which can factor large integers in polynomial time.  This exponential speedup over classical algorithms poses a threat to classical cryptographic systems that rely on the difficulty of factorization. 

\subsubsection*{Unstructured Search} Classical algorithms for unstructured search, such as linear search, require \(O(N)\) time to find a target item in an unsorted database of \(N\) elements. This linear complexity can be prohibitive for large datasets, leading to inefficiencies in data retrieval and processing.

Grover's algorithm provides a quadratic speedup for unstructured search problems, reducing the time complexity to \(O(\sqrt{N})\). This significant improvement allows quantum computers to search through large datasets more efficiently than classical computers, making it a powerful tool for various applications, including database searching and optimization problems.
\subsubsection*{Optimization} Classical optimization techniques, such as gradient descent or genetic algorithms, often struggle with combinatorial problems where the solution space grows exponentially with the number of variables. These methods may require significant computational resources and time to converge to an optimal solution.

Quantum computing also shows promise in optimization problems through techniques such as the Quantum Approximate Optimization Algorithm (QAOA) and quantum annealing. These methods leverage quantum superposition and entanglement to explore solution spaces more effectively than classical optimization algorithms. For example, quantum annealers can tackle combinatorial optimization problems that are classically intractable, such as vehicle routing and portfolio optimization.

 \subsection{Error Tolerance }

Error tolerance is a critical aspect of computing systems, influencing their reliability and performance. Classical computing systems are designed to be robust against errors, while quantum computing systems face unique challenges due to the fragile nature of quantum states. This section compares classical and quantum computing in terms of their error tolerance capabilities.

Classical computing systems are built with error tolerance in mind. They utilize various techniques to detect and correct errors, ensuring that computations can proceed reliably even in the presence of faults. 

Techniques such as parity bits and checksums are used to identify errors in data transmission and storage. These codes help ensure data integrity and enable the system to recover from errors. More advanced techniques, such as Hamming codes and Reed-Solomon codes, allow for the correction of errors without requiring retransmission of data. These codes add redundancy to the information, enabling the recovery of the original data even if some bits are corrupted.

In contrast, quantum computing systems are inherently fragile due to the delicate nature of quantum states. Quantum information is susceptible to errors caused by decoherence, noise, and other environmental factors. This fragility poses significant challenges for the implementation of quantum algorithms and the reliability of quantum computations. To address these challenges, quantum error correction (QEC) techniques have been developed. QEC aims to protect quantum information from errors by encoding logical qubits into entangled states of multiple physical qubits. Some key concepts include: (i) \textit{Stabilizer Codes}: These codes are a class of quantum error correction codes that use stabilizer operators to detect and correct errors without measuring the quantum state directly. The most well-known stabilizer code is the surface code, which is particularly effective for fault-tolerant quantum computation \cite{10.1098/rspa.1998.0166}. (ii) \textit{Fault-Tolerant Quantum Computation}: This approach ensures that quantum operations can be performed reliably even in the presence of errors. It involves designing quantum circuits and algorithms that can tolerate a certain level of noise and errors, allowing for the successful execution of quantum computations \cite{10.1109/aucc.2016.7868220}.

\section{Fundamental Principles of Quantum Mechanics}

Quantum Mechanics (QM) is a branch of Physics that describes the behaviour of matter and energy at the atomic and subatomic levels. It includes a variety of phenomena such as  {\em wave-particle duality}, {\em superposition}, and {\em entanglement}. These phenomena of quantum mechanics are essential for interpreting experimental results and the theoretical underpinnings of Quantum Computing and Quantum Information Systems. Further, the principles of quantum mechanics help us design and implement quantum algorithms and error correction techniques,   secure communication and interdisciplinary research \cite{10.1103/physreva.84.012311}.

The key concepts discussed in this section include wave-particle duality, superposition principle, quantum state, wavefunction, Schrödinger’s equation, Heisenberg uncertainty principle, and the collapse of the wavefunction. These concepts constitute the foundation of quantum mechanics. Further, it provides a framework for understanding the behaviour of matter and energy at the quantum level, influencing theoretical research and practical applications in quantum computing and information systems.

\subsection{Wave-Particle Duality}  
Wave-particle duality illustrates that particles such as electrons and photons exhibit wave-like and particle-like properties. This duality was first articulated through the work of Louis de Broglie, who proposed that all matter has wave characteristics, leading to the development of wave mechanics by Erwin Schrödinger \cite{10.48550/arxiv.0904.4021}. Wave-particle duality is essential for understanding phenomena like interference and diffraction, typically associated with waves, but can also be observed in particles under certain conditions \cite{10.21203/rs.3.rs-475920/v1}. 

Recent studies have further explored this duality, demonstrating its relevance even in macroscopic systems, such as bouncing droplets that mimic quantum behaviour \cite{10.1088/0143-0807/37/5/055706}. The experimental verification of wave-particle duality further explains how quantum systems can be harnessed for information processing \cite{10.1088/1402-4896/ada313}.

From a quantum computing perspective, wave-particle duality is essential for the concept of \textit{quantum superposition}. Qubits can exist in a superposition of states, allowing them to perform multiple calculations simultaneously. This capability is a direct consequence of the wave-like nature of quantum systems, where interference patterns can emerge from the superposition of different quantum states \cite{10.1088/1402-4896/ada313}. 


Moreover, the \textit{entropic uncertainty principle} is closely related to wave-particle duality and plays a significant role in quantum information theory. Research has shown that the uncertainty associated with measuring a quantum system's wave or particle properties can be quantified and is fundamentally linked to the information that can be extracted from the system \cite{10.1038/ncomms6814}. This relationship is crucial for developing quantum communication protocols, where the secure transmission of information relies on the inherent uncertainties of quantum states \cite{10.1038/ncomms6814}. 


The duality principle also supports the design of \textit{quantum interference} experiments, which is exploited in quantum algorithms that utilize interference to amplify correct answers while canceling out incorrect ones \cite{10.1007/s10773-010-0603-z}. 


Furthermore, advancements in experimental techniques, such as the use of \textit{quantum delayed-choice experiments}  illustrate that the decision to measure a quantum system as a wave or a particle can be made after the system has already interacted with its environment, suggesting that quantum information can be processed in ways that defy classical intuitions \cite{10.1126/science.1226719}. This has profound implications for the development of quantum technologies, as it suggests new methods for manipulating and controlling quantum states for computation and communication \cite{10.1038/s41467-021-22887-6}.

\subsection{Superposition Principle} The Superposition Principle states that a quantum system can exist in multiple states simultaneously until it is measured. This principle is crucial for the operation of quantum computers, where qubits can represent both 0 and 1 at the same time, leveraging the power of superposition to perform complex calculations more efficiently than classical computers \cite{10.32474/mams.2020.03.000157}. The superposition of quantum states is foundational to understanding quantum entanglement and interference patterns \cite{10.32474/mams.2020.03.000157}  The implications of the superposition principle are discussed below:
\begin{enumerate}
    \item \textit{Enhanced Computational Power:} In classical computing, a bit can be either 0 or 1, whereas a qubit can exist in a superposition of both states. This property allows quantum computers to explore a vast solution space simultaneously, leading to significant speedups for specific computational tasks. For instance, Shor's algorithm for factoring large integers exploits superposition to achieve exponential speedup over the best-known classical algorithms \cite{10.1103/physrevlett.110.230501}. The ability to create superpositions of qubits allows quantum computers to process information in ways that classical computers cannot, fundamentally changing the landscape of computational capabilities\cite{10.1557/s43577-021-00133-0}. This is a key advantage of quantum computing, allowing for the efficient execution of complex algorithms \cite{10.32474/mams.2020.03.000157}.

    \item \textit{ Quantum Algorithms and Parallelism}: The superposition principle enables quantum algorithms to leverage quantum parallelism.  Algorithms such as Grover's search algorithm provide a quadratic speedup for unstructured search problems by utilising superposition to evaluate multiple possibilities at once \cite{10.1103/physrevlett.110.230501}. 

    \item \textit{ Quantum Information Processing} In quantum information systems, quantum states can be superposed to create complex quantum information protocols, such as quantum teleportation and quantum key distribution. These protocols rely on the ability to create and manipulate superpositions of quantum states to achieve secure communication and efficient information transfer \cite{10.3389/fphy.2024.1456491}. 
\end{enumerate}

Despite these various examples harnessing superposition principles, the 
realizing it at the physical level is challenging. Hence, the experimental realization of superposition states is a significant area of research in quantum computing. Recent studies have focused on creating superpositions of unknown quantum states, which poses challenges due to the inherent uncertainties in quantum measurements \cite{10.1103/physrevlett.116.110403}. The ability to experimentally superpose states is crucial for developing robust quantum computing models and has implications for the scalability of quantum technologies \cite{10.1103/physreva.97.052330}. Understanding the limitations and possibilities of creating superpositions is essential for advancing quantum information science \cite{10.1103/physreva.95.022334}. Also, the concept of superposition has been formalized within the framework of resource theories, which quantify the utility of superposition in quantum information tasks. This approach allows researchers to explore the conditions under which superposition can be harnessed as a resource for computation and communication \cite{10.1103/physrevlett.119.230401}.

\subsection{Quantum State} In quantum computing, state is the basic unit of information, and it can be either $|0\rangle$ or $|1\rangle$, or superposition of both states   expressed mathematically as: 
\[
\psi  = \alpha |0\rangle + \beta |1\rangle
\]

where \( \alpha \) and \( \beta \) are complex coefficients that determine the probability amplitudes of measuring the qubit in either state\cite{10.22533/at.ed.3174152421055}.

The {\em no-cloning theorem }states that it is impossible to create an identical copy of an arbitrary unknown quantum state. This theorem is a direct consequence of the linearity of quantum mechanics.  The {\em no-cloning theorem} is crucial for the security of quantum communication protocols, as it prevents eavesdroppers from perfectly copying quantum information. It ensures that quantum states cannot be duplicated, preserving the integrity of quantum information. This property is fundamental to the development of secure quantum cryptographic systems and protocols \cite{10.1002/que2.45}.

The significance of quantum states from quantum computing and quantum information systems perspectives are discussed below:

\begin{enumerate}

\item  \textit{Quantum Gates and Operations:}
Quantum states are manipulated using quantum gates, which are unitary operations that transform the state of qubits. These operations preserve the quantum nature of the states, allowing for complex computations to be performed through sequences of quantum gates. The ability to create and manipulate quantum states is essential for executing quantum algorithms that leverage the unique properties of quantum mechanics \cite{10.4028/www.scientific.net/amr.267.757}.

\item \textit{Entanglement and Quantum Circuits:}
Quantum states can also be entangled, meaning the state of one qubit is dependent on the state of another, regardless of the distance between them. This phenomenon is crucial for quantum computing, as it enables the creation of quantum circuits that can perform operations on multiple qubits simultaneously \cite{10.1103/physrevlett.101.200501}. Entangled states are used in various quantum algorithms and protocols, enhancing the efficiency and security of quantum information processing \cite{10.1038/nphoton.2013.355}.

\item \textit{Information Encoding and Transmission:}
In quantum information science, quantum states serve as carriers of information. The ability to encode information in quantum states allows for the development of quantum communication protocols, such as quantum key distribution (QKD), which utilizes the principles of quantum mechanics to ensure secure communication. The security of QKD relies on the properties of quantum states, particularly their susceptibility to measurement, which can reveal eavesdropping attempts \cite{10.3389/fphy.2024.1456491}.

\item \textit{Quantum Memory and State Preservation:}
Quantum states can be stored and retrieved using quantum memory systems, which are essential for building scalable quantum networks. The preservation of quantum states during storage is critical for maintaining the integrity of quantum information. Recent advancements in quantum memory technologies have demonstrated the ability to store and retrieve quantum states with high fidelity, enabling the development of robust quantum communication systems \cite{10.1038/ncomms3527}.

\item \textit{Quantum State Tomography:}
Quantum state tomography is a technique used to reconstruct the quantum state of a system based on measurement outcomes. This process is vital for characterizing quantum systems and verifying quantum operations. By performing a series of measurements on a quantum state, researchers can infer the density matrix that describes the state, providing insights into its properties and behavior \cite{10.1088/1367-2630/17/3/033037}. This technique is crucial for experimental implementations of quantum computing and information protocols.

\end{enumerate}
The experimental realization of quantum states, particularly in high-dimensional systems, presents significant challenges. Techniques such as generating squeezed states and manipulating orbital angular momentum (OAM) states have been explored to enhance the capabilities of quantum information systems \cite{10.1007/s11801-008-7153-0}. These advancements enable the creation of complex quantum states that can be used for various applications, including quantum communication and computation.

\subsection{Wavefunction ($\psi$) and Schrödinger’s Equation}  

Wavefunction encapsulates the state of a quantum system. It is a complex-valued function that provides a complete description of the quantum state, allowing for the calculation of probabilities associated with measurement outcomes.   The wavefunction evolves according to the Schrödinger equation, which describes how quantum states change over time.

 Schrödinger equation is a partial differential equation that plays a similar role in quantum mechanics as Newton's laws do in classical mechanics. The equation can be derived from various formulations of quantum mechanics, including Feynman’s path integral formulation. The solutions to the Schrödinger equation provide critical insights into the behaviour of quantum systems, including energy levels and transition probabilities.

 The time-dependent Schrödinger equation is given by:

\begin{equation}
i\hbar \frac{\partial}{\partial t} \Psi(\mathbf{r}, t) = \hat{H} \Psi(\mathbf{r}, t)
\end{equation}

where:
\begin{itemize}
\item \(i\) is the imaginary unit, satisfying \(i^2 = -1\). The imaginary unit, which is essential in quantum mechanics for representing phase information in wavefunctions. It allows for the description of oscillatory behavior typical of quantum states.
\item \(\hbar\) (h-bar) is the reduced Planck's constant, defined as \(\hbar = \frac{h}{2\pi}\), where \(h\) is Planck's constant.
The reduced Planck's constant, a fundamental physical constant that relates the energy of a photon to its frequency. It plays a crucial role in the quantization of physical systems and is a measure of the scale at which quantum effects become significant.

\item \(\Psi(\mathbf{r}, t)\) is the wavefunction of the quantum system, which depends on the position \(\mathbf{r}\) and time \(t\). The wavefunction contains all the information about the system's state. The wavefunction of the quantum system, which encodes all the information about the system's state. The square of the absolute value of the wavefunction, \(|\Psi(\mathbf{r}, t)|^2\), gives the probability density of finding a particle at position \(\mathbf{r}\) at time \(t\). The wavefunction is a complex function, and its phase and amplitude contain important physical information about the system.

\item \(\hat{H}\) is the Hamiltonian operator, which represents the total energy of the system (kinetic and potential energy). It is an operator that acts on the wavefunction. The Hamiltonian operator, which represents the total energy of the quantum system. It includes both kinetic energy (associated with the motion of particles) and potential energy (associated with the forces acting on particles). The Hamiltonian is a key operator in quantum mechanics, and its eigenvalues correspond to the possible energy levels of the system.
\end{itemize}

This evolution is governed by the principles of quantum mechanics, allowing for complex operations to be performed on qubits. The ability to manipulate wavefunctions through quantum gates is essential for implementing quantum algorithms, such as the Variational Quantum Eigensolver (VQE), which relies on approximating the ground-state wavefunction to compute eigenvalues \cite{10.1088/1367-2630/17/3/033037}.

\subsection{Heisenberg Uncertainty Principle and Quantum Measurement Problem} The  Heisenberg Uncertainty Principle articulates a fundamental limit to the precision with which pairs of physical properties, such as position and momentum, can be known simultaneously. This principle is not merely a statement about measurement limitations but reflects intrinsic properties of quantum systems \cite{10.1142/10541}. It leads to the concept of the {\em quantum measurement problem}, where the act of measurement affects the system being observed \cite{10.1142/10541}.  The {\em quantum measurement problem}  raises questions about the nature of reality, the role of the observer, and the transition from quantum superpositions to classical outcomes. 

Quantum measurement problem, which can be chareterized as follows, arises from the apparent conflict between the linear evolution of quantum states, described by the Schrödinger equation, and the discrete outcomes observed during measurements.  
\begin{enumerate}
    \item \textit{ Superposition and Measurement}: Quantum systems can exist in superpositions of states, where a particle can be in multiple states simultaneously. However, when a measurement is made, the system appears to {\em collapse} to a single outcome. This collapse is not described by the deterministic evolution of the wavefunction, leading to questions about how and why this transition occurs \cite{10.1238/physica.regular.072a00290}.

\item \textit{Observer Effect}: The act of measurement in quantum mechanics seems to affect the system being observed. This raises philosophical questions about the role of the observer and whether the observer's knowledge influences the outcome of a measurement. The measurement problem challenges classical intuitions about reality and determinism, suggesting that at a fundamental level, quantum systems behave differently from classical systems \cite{10.1016/j.entcs.2006.12.012}.

\end{enumerate}

The {\em quantum measurement problem} has several implications for quantum computing. These are discussed below:

\begin{enumerate}
    \item \textit{Measurement-Based Quantum Computing}: In measurement-based quantum computing (MBQC), quantum information is processed through a series of measurements on entangled states. The one-way quantum computer model, introduced by Raussendorf and Briegel, relies on performing projective measurements on a highly entangled resource state, such as a cluster state \cite{10.1142/s0219749909005699}. Understanding the measurement problem is essential for developing robust MBQC protocols and ensuring that the desired quantum operations are achieved through measurements.

\item \textit{Error Correction and Fault Tolerance}: The measurement problem highlights the importance of error correction in quantum computing. Quantum systems are susceptible to decoherence and noise, which can affect measurement outcomes. Developing error correction techniques that account for the measurement problem is crucial for building reliable quantum computers capable of performing complex computations \cite{10.1117/12.620302}.

\item \textit{Quantum Algorithms}: The measurement problem influences the design and implementation of quantum algorithms. For instance, algorithms that rely on measurements must consider the effects of measurement-induced collapse and the potential loss of quantum coherence. Understanding the nuances of the measurement process is vital for optimizing quantum algorithms and ensuring their success \cite{10.1103/physreva.75.012337}.

\item \textit{Quantum Cryptography}: In quantum key distribution (QKD), the security of the protocol relies on the properties of quantum measurements. The measurement problem raises questions about the nature of information and how it can be securely transmitted. Understanding the measurement process is essential for developing secure quantum communication protocols that can withstand eavesdropping attempts \cite{10.36347/sjpms.2020.v07i08.001}.

\item \textit{State Verification and Characterization}: Quantum state tomography, a technique used to reconstruct the wavefunction of a quantum system, is closely related to the measurement problem. The accuracy of state reconstruction depends on the measurement process and the ability to extract meaningful information from quantum states. Addressing the measurement problem is crucial for improving the reliability of quantum state verification techniques \cite{10.1364/fio.2017.fth3e.6}.

\end{enumerate}
Several attempts have been made to investigate the nature of measurement enhancing our understanding of quantum phenomena \cite{10.3389/fphy.2020.00139} and its implications for advancements in quantum information science.  To resolve the measurement problem, various interpretations, including the Copenhagen interpretation, Many-World interpretation, and objective collapse, have been formulated.    Each interpretation offers a different perspective on the nature of quantum states and the measurement process, reflecting the ongoing debate within the physics community \cite{10.3389/fphy.2020.00139}.

 \section{Quantum Entanglement and Nonlocality}
Quantum entanglement and nonlocality are phenomena that challenge classical intuitions about separability and locality, providing unique opportunities for quantum technologies. These concepts are described below, considering their significance and applications in quantum computing and quantum information science.

\subsection{Quantum Entanglement} 
Quantum entanglement is a phenomenon where two or more quantum particles become correlated so that the state of one particle cannot be described independently of the state of the other(s), regardless of the distance separating them. This correlation persists even when large distances separate the entangled particles. 

Entangled states are integral to the functioning of quantum gates, such as the Controlled-NOT (CNOT) gate, which creates entanglement between qubits. Quantum circuits often rely on entangled states to perform complex computations efficiently \cite{10.3390/photonics9020058}.

Entanglement is also significant for many quantum algorithms, enabling impossible operations in classical computing. For instance, entangled states are used in quantum teleportation, where the quantum state of a particle is transmitted from one location to another without physically transferring the particle itself \cite{10.1088/0953-4075/44/6/065504}.

In measurement-based quantum computing (MBQC), entangled states, such as cluster states, serve as resources for computation. The entire computation is performed through adaptive measurements on these entangled states, showcasing the power of entanglement in quantum processing \cite{10.1088/1751-8121/ac3bea}.

\subsection{Nonlocality} 
The {\em nonlocality} is one of the phenomena closely related to entanglement.   It states that measurement outcomes of entangled particles are correlated in a way that classical physics cannot explain. Further, Bell's theorem explaining the reason behind entanglement demonstrates that no local hidden variable theory can account for the correlations observed in entangled systems, implying that the results of measurements on one particle can instantaneously affect the state of another, regardless of the distance between them \cite{10.1038/ncomms3828}. Bell's theorem has challenged classical notions of locality suggesting that the universe may be fundamentally nonlocal. This understanding is crucial for developing quantum technologies, as it informs the design of quantum algorithms and protocols that leverage entanglement and nonlocality \cite{10.1103/physreva.82.062101}.

Various quantum algorithms have been developed based on the phenomenon of nonlocality. For example, nonlocality underpins the security of quantum cryptographic protocols, such as quantum key distribution (QKD). The correlations established by entangled particles ensure that any attempt at eavesdropping can be detected, as it would disturb the entangled state and reveal the intruder's presence \cite{10.48550/arxiv.2003.01011}.

Nonlocality allows for device-independent quantum information protocols whose security does not rely on trusting the devices used. This is particularly important when the devices may be compromised, or untrusted \cite{10.1142/s0219749914500300}.

Nonlocality serves as a tool for testing the foundations of quantum mechanics. Experiments designed to test Bell's inequalities provide insights into the nature of quantum correlations and the validity of quantum mechanics as a complete theory \cite{10.48550/arxiv.2003.01011}.

While Bell's theorem highlights the limitations of classical Physics, the Einstein-Podolsky-Rosen (EPR) paradox does the same for Quantum Mechanics. The EPR paradox is a thought experiment that questions the completeness of quantum mechanics by suggesting that if quantum mechanics is correct, entangled particles exhibit correlations that local hidden variables cannot explain. The EPR paradox has stimulated extensive discussions about the interpretation of quantum mechanics and the nature of quantum states. It has led to the development of various interpretations, including the Many-Worlds interpretation and the Copenhagen interpretation. Understanding the implications of the EPR paradox is essential for researchers entering the field of quantum computing, as it lays the groundwork for the philosophical considerations of quantum information science.
\begin{table}[!ht]
    \centering
    \begin{tabular}{|p{1.1in}|p{0.9in}|p{0.7in}|p{0.9in}|p{1in}|p{1in}|}
       
       \hline 
       \textbf{Interpretation}	& \textbf{Wavefunction Reality}	& \textbf{Collapse?}	& \textbf{Deterministic?}	& \textbf{Role of Observer?}	& \textbf{Key Features} \\ \hline 
       
Copenhagen Interpretation	& Mathematical tool only	& Yes, at measurement	& No.  	& Essential and causes collapse	& Standard interpretation, no hidden variables \\ \hline 

Many-Worlds Interpretation 	& Real  	& No collapse. All outcomes happen.	& Yes  	 & No special role	& All possibilities occur in separate branches \\ \hline 

Objective Collapse Theories & Real	& Yes, spontaneously	&No 	& Observer not required	& Wavefunction collapses due to an intrinsic process \\ \hline 

Bohmian Mechanics	& Real and pilot wave guides particles	& No collapse	& Yes,  hidden variables determine outcomes	& No special role	& Deterministic, nonlocal hidden variable theory  \\ \hline 
QBism 	& Subjective knowledge& 	Yes, but personal&	No 	&Observer’s belief updates	& Treats quantum mechanics as a tool for updating knowledge \\ \hline 
    \end{tabular}
    \caption{A Comparison of Various Philosophical Interpretations}
    \label{inter}
\end{table}

\section{Quantum Measurement and Interpretation}
The quantum measurement problem has led to various philosophical interpretations that attempt to explain the nature of quantum mechanics, particularly regarding how measurements affect quantum systems. Each interpretation provides a different perspective on the implications of quantum measurement, entanglement, and the nature of reality. 

\subsection{Copenhagen Interpretation}
The Copenhagen interpretation, primarily associated with Niels Bohr and Werner Heisenberg, posits that quantum mechanics is inherently probabilistic. According to this view, a quantum system exists in a superposition of states until a measurement is made, at which point the wavefunction collapses to a definite state.

The Copenhagen interpretation emphasizes the role of the observer in defining the properties of a quantum system. This perspective raises questions about the nature of reality and whether quantum states exist independently of observation \cite{10.4236/jqis.2011.12005}.

For software engineers and researchers in quantum computing, understanding the Copenhagen interpretation is essential for grasping the probabilistic nature of quantum algorithms and the implications of measurement in quantum information processing \cite{10.1007/s10702-006-1009-2}.

\subsection{Many-Worlds Interpretation}

Proposed by Hugh Everett III, the Many-Worlds interpretation suggests that every quantum event leads to a branching of the universe into multiple, non-communicating parallel realities. In this view, all possible outcomes of a quantum measurement occur, each in its own separate branch of the universe.

The Many-Worlds interpretation provides a deterministic framework for quantum mechanics, eliminating the need for wavefunction collapse. This perspective can be appealing for those seeking a coherent understanding of quantum phenomena without invoking randomness \cite{10.1007/s10773-009-0239-z}.

In the context of quantum computing, the Many-Worlds interpretation suggests that quantum algorithms could explore multiple computational paths simultaneously, potentially leading to enhanced computational power. Understanding this interpretation can help researchers conceptualize the implications of quantum superposition and entanglement in algorithm design.

\subsection{Bohmian Mechanics}

Bohmian mechanics, also known as pilot-wave theory, is a hidden-variable theory that provides a deterministic view of quantum mechanics. It posits that particles have definite positions and velocities, guided by a "pilot wave" described by the wavefunction.

Bohmian mechanics offers a clear alternative to the probabilistic nature of standard quantum mechanics. This interpretation allows for a more intuitive understanding of quantum phenomena, as it reintroduces determinism into the description of quantum systems \cite{10.1016/j.physleta.2008.09.053}.
The deterministic nature of Bohmian mechanics can inform the development of quantum algorithms and protocols, particularly in scenarios where understanding particle trajectories is essential \cite{10.1007/978-3-642-30690-7_2}. Researchers can explore how Bohmian mechanics might provide insights into the behavior of quantum systems in computational contexts.

\subsection{Objective Collapse Theories}

Objective collapse theories propose that wavefunction collapse is a physical process that occurs independently of observation. These theories suggest that certain mechanisms cause the wavefunction to collapse when a system reaches a specific threshold or interacts with the environment.

Objective collapse theories provide a framework for understanding how measurements affect quantum systems without relying on the observer's role. This perspective can help clarify the nature of quantum measurements and their implications for quantum information processing \cite{10.1126/science.1501466}.

Understanding objective collapse theories can inform the design of quantum technologies, particularly in addressing issues related to decoherence and the stability of quantum states in practical applications \cite{10.1007/s10701-006-9065-9}.

\subsection{Quantum Bayesianism (QBism)}

Quantum Bayesianism, or QBism, is a subjective interpretation of quantum mechanics that emphasizes the role of the observer's knowledge and beliefs in the measurement process. According to QBism, quantum probabilities represent an observer's subjective degrees of belief about the outcomes of measurements rather than objective properties of the system.

QBism challenges traditional views of quantum states as objective entities, suggesting that they are instead tools for making predictions about measurement outcomes. This perspective can influence how researchers approach quantum algorithms and their interpretations \cite{10.1111/0591-2385.00293}.

Understanding QBism can provide insights into the nature of quantum information and its processing. It emphasizes the importance of the observer's perspective in quantum information systems, which can inform the development of quantum communication protocols and decision-making processes in quantum computing \cite{10.1007/s10702-005-1128-1}.

Table \ref{inter} compares all the interpretations that have compared in this section.

\section{Quantum Operators and Observables}
 Quantum operators and observables are essential for understanding how measurements are made in quantum systems and how these measurements relate to physical observables. The key concepts related to quantum operators and observables, including operators, commutators and non-commutativity, Hermitian operators, and expectation values are discussed below.

\subsection{Operators} In quantum mechanics, operators are mathematical entities that act on the state vectors in a Hilbert space. They are used to represent physical observables, such as position, momentum, and energy. An operator \( \hat{A} \) acts on a state vector \( |\psi\rangle \) to produce another state vector:

\[
\hat{A} |\psi\rangle = |\phi\rangle
\]

where \( |\phi\rangle \) is the resulting state after the operator \( \hat{A} \) has acted on \( |\psi\rangle \).

 Operators are central to the formulation of quantum mechanics. They allow for the mathematical representation of physical quantities and facilitate the computation of measurement outcomes. The properties of operators, such as their eigenvalues and eigenvectors, provide insights into the possible results of measurements and the behavior of quantum systems.

\subsection{Commutators and Non-Commutativity} The commutator of two operators \( \hat{A} \) and \( \hat{B} \) is defined as:

\[
\hat{A}, \hat{B} = \hat{A}\hat{B} - \hat{B}\hat{A}
\]

If the commutator is zero, the operators are said to commute; if it is non-zero, they do not commute.  Non-commutativity implies that certain pairs of observables cannot be measured simultaneously with arbitrary precision, as articulated by the Heisenberg uncertainty principle. For example, the position and momentum operators satisfy the commutation relation:

\[
\hat{x}, \hat{p} = i\hbar
\]

This non-commutativity results in inherent uncertainties in measurements and is a key feature that differentiates quantum mechanics from classical mechanics.

 \subsection{Hermitian Operators}
An operator \( \hat{A} \) is said to be Hermitian (or self-adjoint) if it satisfies the condition:

\[
\hat{A} = \hat{A}^\dagger
\]

where \( \hat{A}^\dagger \) is the adjoint of \( \hat{A} \). Hermitian operators have real eigenvalues and orthogonal eigenvectors.

Hermitian operators represent measurable quantities. The eigenvalues of a Hermitian operator correspond to the possible outcomes of the observable measurement, while the eigenvectors indicate the states associated with those outcomes. The requirement for observables to be represented by Hermitian operators ensures that measurement results are real and physically meaningful.

\subsection{Expectation Value}The expectation value of an observable represented by a Hermitian operator \( \hat{A} \) in a quantum state \( |\psi\rangle \) is given by:

\[
\langle A \rangle = \langle \psi | \hat{A} | \psi \rangle
\]

This expression calculates the average value of the observable when measurements are made on a large number of identically prepared systems in the state \( |\psi\rangle \).

 The expectation value provides a statistical measure of the observable's value in a given quantum state. It is a key concept in quantum mechanics, as it allows for the prediction of measurement outcomes and the characterization of quantum states. The expectation value plays a vital role in quantum computing and information science, where it is used to evaluate the performance of quantum algorithms and the behavior of quantum systems under various operations.

 \section{Quantum Statistics and Particles} Quantum statistics describes how indistinguishable particles behave under various conditions. Their statistical properties govern the behaviour of particles, which are categorized into two main types: {\em fermions} and {\em bosons}. This section elaborates on key concepts in quantum statistics, including the Pauli Exclusion Principle, Bose-Einstein statistics, Fermi-Dirac statistics, and their implications for quantum computing and quantum information systems.

\subsection{ Pauli Exclusion Principle}

The Pauli Exclusion Principle states that no identical fermions can simultaneously occupy the same quantum state. This principle is a direct consequence of the antisymmetry of the wavefunction for fermions, which are particles with half-integer spin (e.g., electrons, protons, and neutrons).

The Pauli Exclusion Principle leads to the unique behavior of fermions, resulting in the formation of electron shells in atoms and the stability of matter. It explains why electrons occupy atomic orbitals in a specific order, contributing to the periodic table of elements \\cite{10.3934/krm.2019014}

 In quantum computing, the Pauli Exclusion Principle is essential for understanding the behavior of qubits representing fermionic systems. It impacts the design of quantum algorithms and error correction methods that depend on the properties of fermionic states \cite{10.1088/1742-6596/845/1/012030}.

\subsection{ Bose-Einstein Statistics}Bose-Einstein statistics govern the behavior of bosons, which are particles with integer spin (e.g., photons, gluons, and helium-4 atoms). Unlike fermions, bosons can occupy the same quantum state, leading to phenomena such as Bose-Einstein condensation.

At extremely low temperatures, a group of bosons can occupy the same ground state, resulting in a macroscopic quantum state known as a Bose-Einstein condensate. This phenomenon has implications for quantum computing, as it enables the exploration of new quantum states and the development of novel quantum technologies \cite{10.1007/s10701-022-00596-4}.

 Bose-Einstein statistics are significant in quantum information systems, especially concerning quantum communication and quantum optics, where manipulating bosonic states is crucial for effective information transfer \cite{10.1021/acsami.7b09805}.

\subsection{ Fermi-Dirac Statistics}

Fermi-Dirac statistics describe the distribution of indistinguishable fermions over energy states in a system. The Fermi-Dirac distribution function gives the probability of occupancy of energy states by fermions at a given temperature.

Fermi-Dirac statistics are crucial for understanding the behavior of electrons in solids, particularly in the context of semiconductors and metals. The distribution function helps explain phenomena such as electrical conductivity and heat capacity in materials \cite{10.1007/s10773-014-2140-7}.

 In quantum computing, Fermi-Dirac statistics inform the design of quantum algorithms that involve fermionic systems, such as quantum simulations of electronic structures and materials. Understanding the statistical behavior of fermions is essential for developing efficient quantum algorithms and error correction techniques \cite{10.1088/1742-6596/845/1/012030}.

\subsection{Eigenvalues and Eigenvectors in Quantum Statistics}

In the context of quantum statistics, observables are represented by operators, and the eigenvalues of these operators correspond to the measurable quantities associated with the particles. The eigenvectors represent the quantum states associated with those eigenvalues.

 The eigenvalues and eigenvectors provide a framework for understanding measurement outcomes in quantum systems. When a measurement is made, the system collapses into one of the eigenstates, and the corresponding eigenvalue is observed \cite{10.3934/krm.2019014}.

 Eigenvalues and eigenvectors are essential for characterizing quantum states in both fermionic and bosonic systems. They help in the analysis of quantum systems and the development of quantum algorithms that rely on the manipulation of these states \cite{10.48550/arxiv.2002.10133}.

\section{Representational Theories for Quantum Phenomenon}
 Quantum states are essential components of quantum systems and describe their behaviour. They can be represented in several mathematical forms, including state vectors and density matrices, and are analyzed within the framework of Hilbert spaces. The following discussion will elaborate on these concepts and related topics.

A {\em Hilbert space} is a complete vector space equipped with an inner product that allows for measuring angles and distances between vectors.  Hilbert space provides the framework for defining quantum states, operators, and observables. The completeness of Hilbert space ensures that limits of sequences of vectors, i.e. states, are also contained within the space, which is essential for the mathematical rigour of quantum mechanics \cite{10.48550/arxiv.2103.05301}.

In quantum mechanics, the states of a quantum system are represented as vectors in a Hilbert space.  A {\em state vector} represents a quantum state in a Hilbert space \cite{10.1088/0305-4470/35/3/315}. For a single qubit, the state vector can be expressed as:

\[|\psi\rangle = \alpha |0\rangle + \beta |1\rangle \]

where \( |0\rangle \) and \( |1\rangle \) are the basis states, and \( \alpha \) and \( \beta \) are complex coefficients satisfying the normalization condition \( |\alpha|^2 + |\beta|^2 = 1 \). They allow for calculating probabilities of measurement outcomes and the evolution of states under quantum operations. 

\subsubsection*{Example 1: Representing A Qubit in Superposition as a state vector} Consider a single qubit in the Hadamard state: 
\[ \ket{\psi} = \frac{1}{\sqrt{2}} \ket{0} +  \frac{1}{\sqrt{2}} \ket{1}\]
In vector form, this is written as:
\begin{math}
    \ket{\psi} = \frac{1}{\sqrt{2}}\begin{pmatrix}
        1\\
        1\\ 
    \end{pmatrix}
\end{math}
If we measure this state in the standard basis $\{\ket{0},\ket{1}\}$, the probabilities are: $P(\ket{0}) = \lvert\frac{1}{\sqrt{2}}\rvert^2 = \frac{1}{2}$ and $P(\ket{1}) = \lvert\frac{1}{\sqrt{2}}\rvert^2 = \frac{1}{2}$. Thus, a measurement collapses the state to either $\ket{0}$ or $\ket{1}$ with equal probability.

A {\em single state vector} in Hilbert space represents a {\em pure state} and can be entirely described by its wave function. Pure states exhibit maximum knowledge about the system. A {\em density matrix} represents a {\em mixed state} and arises when there is uncertainty about the system's state, often due to environmental interactions. Mixed states can be thought of as statistical mixtures of pure states.

The {\em density matrix or density operator} is a mathematical representation that describes the statistical state of a quantum system.  For a mixed state, the density matrix can be expressed as a weighted sum of projectors onto the pure states:

\[
\rho = \sum_i p_i |\psi_i\rangle \langle \psi_i|
\]

where \( p_i \) are the probabilities of the system being in the state \( |\psi_i\rangle \).

For a pure state, the density matrix is given by:

\[
\rho = |\psi\rangle \langle \psi|
\]

\subsubsection*{Example 2: Representing a pure state using Density Matrix} The density matrix for a pure state 
\[ \rho = \ket{\psi}\bra{\psi}\] For our Hadamard state:
\begin{center}
    \begin{math}
        \rho= \left(\frac{1}{\sqrt{2}}\begin{pmatrix}
            1\\
            1\\
        \end{pmatrix} \right) \left(\frac{1}{\sqrt{2}}\begin{pmatrix}
            1 &   1\\
        \end{pmatrix} \right) = \frac{1}{2}\begin{pmatrix}
            1 & 1\\
            1 & 1\\
        \end{pmatrix}
    \end{math}
\end{center}

While interpreting the density matrix
the diagonal elements, $(\rho_{00},\rho_{11})$ represent the probabilities of measuring  $\ket{0}$ or $\ket{1}$. The off-diagonal elements $(\rho_{01},\rho_{10})$ encode coherence (quantum interference effects).
If the system undergoes decoherence, the off-diagonal terms vanish, transforming it into a classical mixed state.

\subsubsection*{Example 3: Representing a mixed state  using Density Matrix}
A mixed state represents a statistical ensemble of multiple quantum states
\[ \rho = p_{1} \ket{\psi_{1}} \bra{\psi_{1}} +  p_{2} \ket{\psi_{2}} \bra{\psi_{2}}\]

For example, if a qubit is in $\ket{0}$ with 50\% probability and $\ket{1}$  with 50\% probability, its density matrix is:

         \[\rho = \frac{1}{2}\ket{0}\bra{0} + \frac{1}{2}\ket{1}\bra{1}\] 

         \[
         \rho = \frac{1}{2}\begin{pmatrix}
             1 & 0 \\
             0 & 0 \\
         \end{pmatrix} +  \frac{1}{2}\begin{pmatrix}
             0 & 0 \\
             0 & 1 \\
         \end{pmatrix}\] 
         \[
         \rho =\begin{pmatrix}
             0.5 & 0 \\ 
             0  & 0.5 \\
         \end{pmatrix}\]

 Unlike a pure state, this matrix has no off-diagonal elements, meaning there is no coherence or superposition—just a classical probability distribution over states.

Unlike Hilbert space, the density matrix provides a complete quantum system description, including pure and mixed states. It is imperative in quantum information science, as it allows for characterising quantum states that are not pure due to interactions with the environment (decoherence) \cite{10.1088/0305-4470/35/25/305}. The density matrix formalism is essential for quantum computing, especially in scenarios involving noise and mixed states \cite{10.1140/epjd/e2017-70752-3}.

The distinction between pure and mixed states is crucial for quantum computing and information processing. Pure states are ideal for quantum algorithms, while mixed states represent more realistic scenarios in practical quantum systems where decoherence and noise are present. Understanding how to manipulate and measure both types of states is essential for developing robust quantum technologies \cite{10.1088/0305-4470/35/3/315}. Hence, Table \ref{density} compares both the Hilbert Space and Density Matrix as representational theories.

\begin{table}[!ht]
    \centering
    \begin{tabular}{|p{1.5in}|p{2in}|p{2in}|}
    \hline 
       \textbf{Feature}	& \textbf{Hilbert Space Approach}	&  \textbf{Density Matrix Approach} \\ \hline 
What It Describes	& Quantum states as vectors in an abstract space.&	Quantum states as statistical mixtures in operator form. \\ \hline 
Physical \newline Interpretation	& Represents an isolated, ideal quantum system. &	Represents realistic quantum systems, including decoherence.  \\ \hline 
Handling of Mixed States&	Cannot describe mixed states.&	Naturally describes both pure and mixed states.  \\ \hline 
Computational \newline Complexity &	Computationally efficient for small systems (state vectors scale as $2^n$ & 
More computationally expensive (matrices scale as $2^n X 2^n$ \\ \hline 
Relevance in Quantum Computing	& Used in quantum algorithms, quantum circuits. &	Used in quantum noise models, error correction. \\ \hline 
    \end{tabular}
    \caption{Comparing Hilbert Space and Density Matirx representation}
    \label{density}
\end{table}

In quantum mechanics, observables i.e. such as position, momentum, and spin are represented by Hermitian operators. The {\em eigenvalues} of these operators correspond to the possible measurement outcomes, while the {\em eigenvectors} represent the states associated with those outcomes.

The eigenvalue equation for an operator \( \hat{A} \) is given by:

\[ 
\hat{A} |\phi_n\rangle = a_n |\phi_n\rangle
\]

where \( a_n \) are the eigenvalues and \( |\phi_n\rangle \) are the corresponding eigenvectors. The measurement of an observable yields one of the eigenvalues, and the system collapses into the corresponding eigenstate. This concept is fundamental for understanding quantum measurements and the probabilistic nature of quantum mechanics \cite{10.1088/0031-8949/79/06/065013}.

 Other than Hilbert Space and Density Matrix, the following are some of the other mathematical formalisms that support the analysis of Quantum Systems from a Computing and information Science perspective.
 
\begin{enumerate}

    \item \textbf{Information Theory and Quantum Mechanics}

The relationship between information theory and quantum mechanics has been a significant area of research, particularly in the context of quantum information systems. Classical information theory, as developed by Shannon, provides a framework for understanding data transmission and processing. However, quantum information theory extends these concepts to account for the unique properties of quantum systems, such as superposition and entanglement \cite{10.1007/s00453-002-0971-8}. This extension has led to the development of quantum cryptography and secure communication protocols, which leverage the principles of quantum mechanics to enhance information security. 
\item \textbf{ Contextuality and Quantum Computation}

Contextuality refers to the idea that the outcome of a measurement cannot be understood independently of the context in which it is made. This concept is crucial for understanding quantum computation, as it underlies phenomena such as quantum entanglement and nonlocality\cite{10.1038/nature13460}. The role of contextuality in quantum computation has been explored in various studies, suggesting that it achieves advantages in quantum information processing \cite{10.1038/nature13460} and fault-tolerant quantum algorithms and protocols.

\item \textbf{Operator Algebras and Quantum Foundations}

Operator algebras, particularly $W^{*}$-algebras, provide a mathematical framework for understanding the foundations of quantum mechanics and its applications in quantum computing. These algebras allow for a rigorous treatment of quantum observables and measurements, facilitating the analysis of quantum systems \cite{10.48550/arxiv.1510.06649}. The connection between operator algebras and quantum computation has been explored in various contexts, including the semantics and verification of quantum algorithms.  

\item \textbf{ The Role of Category Theory}

Category theory has emerged as a powerful tool in the foundations of quantum mechanics and quantum computing. It provides a high-level abstraction that can unify various mathematical structures and concepts in quantum theory. For instance, category theory has been used to analyze quantum processes and the relationships between different quantum systems \cite{10.32388/qj8nvr.2}. This approach allows for a more generalized understanding of quantum phenomena and their implications for computation and information processing.
\end{enumerate}

\begin{figure}
    \centering
    \includegraphics[width=0.9\linewidth]{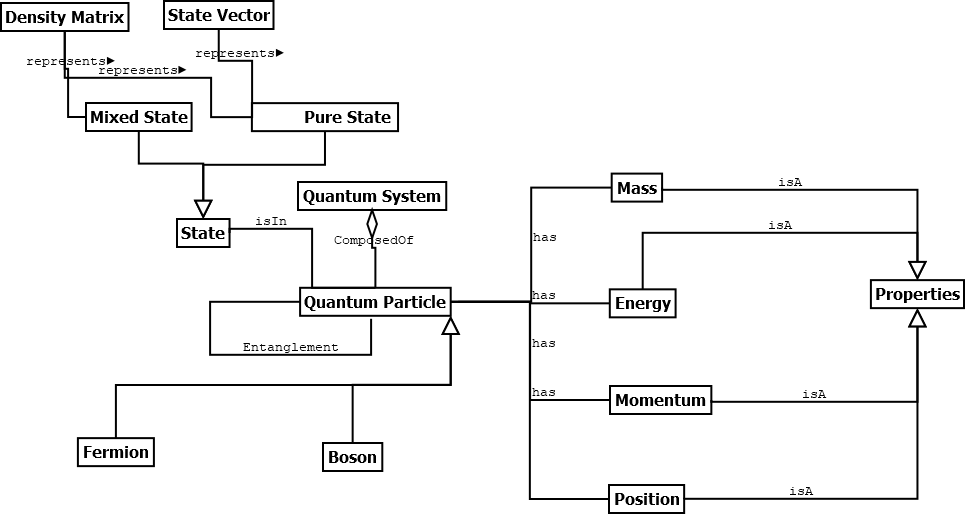}
    \caption{A Conceptual Model capturing Key Concepts}
    \label{cm}
\end{figure}
\section{A Conceptual Model for Quantum concepts}The key concepts of quantum mechanics can be illustrated in a conceptual model, as shown in Figure \ref{cm}. This model uses UML notations and effectively captures the following relationships:
\begin{enumerate}
    \item  A quantum system consists of quantum particles.
\item  Quantum particles can be classified as either fermions or bosons.
\item  Quantum particles possess properties such as mass, momentum, energy, and position.
\item  Each quantum particle is associated with a state.
\item  A wavefunction describes this state.
\item  A quantum state can be categorized as either a pure state or a mixed state.
\item  States can be represented as vectors in Hilbert space or as density matrices.
\item  Quantum particles may exhibit correlations or associations through entanglement.
\end{enumerate}
\section{Conclusions}
Quantum mechanics is the foundation of quantum computing and information science, making its understanding essential for software engineers adopting quantum technologies. This paper provides a primer on fundamental quantum mechanics concepts, including wave-particle duality, superposition, entanglement, quantum measurement, and quantum operators. By presenting these principles from a computational perspective, the paper aims to bridge the knowledge gap between classical computing and quantum computing, equipping software engineers with the theoretical background necessary to engage with emerging quantum technologies. The paper comprehensively covers fundamental quantum concepts relevant to computing and makes these complex topics more accessible to software engineers. Further, it establishes a clear connection between quantum mechanics and computing – the discussion of quantum states, operators, and measurement principles helps software engineers understand the foundation upon which quantum computation is built.

\bibliographystyle{plain}

\begin{thebibliography}{10}

\bibitem{10.1007/s10773-014-2140-7}
A.~A. Abutaleb.
\newblock Unified statistical distribution of quantum particles and symmetry.
\newblock {\em International Journal of Theoretical Physics}, 53:3893--3900, 2014.

\bibitem{10.3934/krm.2019014}
L.~Arkeryd and A.~Nouri.
\newblock On a boltzmann equation for haldane statistics.
\newblock {\em Kinetic \&Amp; Related Models}, 12:323--346, 2019.

\bibitem{10.47852/bonviewjdsis3202656}
M.~W. Arshad, I.~Murtza, and M.~A. Arshad.
\newblock Applications of quantum computing in health sector.
\newblock {\em Journal of Data Science and Intelligent Systems}, 1:19--24, 2023.

\bibitem{10.1103/physreva.82.062101}
J.~A.~i. Batle and M.~Casas.
\newblock Nonlocality and entanglement in thexymodel.
\newblock {\em Physical Review A}, 82, 2010.

\bibitem{10.1103/physrevlett.110.230501}
X.~Cai, C.~Weedbrook, Z.~Su, M.~Chen, M.~Gu, M.~Zhu, L.~Li, N.~Liu, C.~Lu, and J.~Pan.
\newblock Experimental quantum computing to solve systems of linear equations.
\newblock {\em Physical Review Letters}, 110, 2013.

\bibitem{10.1103/physreva.85.032326}
A.~M. Chen, S.~Y. Cho, and M.~D. Kim.
\newblock Implementation of a three-qubit toffoli gate in a single step.
\newblock {\em Physical Review A}, 85, 2012.

\bibitem{10.1038/s41467-021-22887-6}
X.~Chen, Y.~Deng, S.~Liu, T.~Pramanik, J.~Mao, J.~Bao, C.~Zhai, T.~Dai, H.~Yuan, J.~Guo, S.~Fei, M.~Huber, B.~Tang, Y.~Yang, Z.~Li, Q.~He, Q.~Gong, and J.~Wang.
\newblock A generalized multipath delayed-choice experiment on a large-scale quantum nanophotonic chip.
\newblock {\em Nature Communications}, 12, 2021.

\bibitem{10.1103/physreva.84.012311}
Giulio Chiribella, Giacomo~Mauro D’Ariano, and Paolo Perinotti.
\newblock Informational derivation of quantum theory.
\newblock {\em Physical Review A}, 2011.

\bibitem{10.1038/ncomms6814}
P.~J. Coles, J.~Kaniewski, and S.~Wehner.
\newblock Equivalence of wave–particle duality to entropic uncertainty.
\newblock {\em Nature Communications}, 5, 2014.

\bibitem{10.1007/s10773-009-0239-z}
E.~Conte.
\newblock A reformulation of von neumann’s postulates on quantum measurement by using two theorems in clifford algebra.
\newblock {\em International Journal of Theoretical Physics}, 49:587--614, 2010.

\bibitem{10.1016/j.entcs.2006.12.012}
V.~Danos, E.~D’Hondt, E.~Kashefi, and P.~Panangaden.
\newblock Distributed measurement-based quantum computation.
\newblock {\em Electronic Notes in Theoretical Computer Science}, 170:73--94, 2007.

\bibitem{10.1142/s0219749914500300}
A.~Daskin, A.~Grama, and S.~Kais.
\newblock Quantum random state generation with predefined entanglement constraint.
\newblock {\em International Journal of Quantum Information}, 12:1450030, 2014.

\bibitem{10.1038/ncomms3527}
D.~Ding, Z.~Zhou, B.~Shi, and G.~Guo.
\newblock Single-photon-level quantum image memory based on cold atomic ensembles.
\newblock {\em Nature Communications}, 4, 2013.

\bibitem{10.1103/physreva.97.052330}
S.~Dogra, G.~Thomas, S.~Ghosh, and D.~Suter.
\newblock Superposing pure quantum states with partial prior information.
\newblock {\em Physical Review A}, 97, 2018.

\bibitem{10.1238/physica.regular.072a00290}
D.~Dragoman.
\newblock Phase space formulation of quantum mechanics. insight into the measurement problem.
\newblock {\em Physica Scripta}, 72:290--296, 2005.

\bibitem{10.1007/978-3-642-30690-7_2}
D.Durr, S.Goldstein, and N.Zanghı.
\newblock Quantum equilibrium and the origin of absolute uncertainty.
\newblock {\em Quantum Physics Without Quantum Philosophy}, pages 23--77, 2012.

\bibitem{10.1140/epjd/e2017-70752-3}
A.~Frydryszak, M.~I. Samar, and V.~M. Tkachuk.
\newblock Quantifying geometric measure of entanglement by mean value of spin and spin correlations with application to physical systems.
\newblock {\em The European Physical Journal D}, 71, 2017.

\bibitem{10.1007/s10701-022-00596-4}
J.~C. Garrison.
\newblock Quantum statistics of identical particles.
\newblock {\em Foundations of Physics}, 52, 2022.

\bibitem{10.1088/0953-4075/44/6/065504}
J.~Guo, Y.~Mi, J.~Zhang, and H.~Song.
\newblock Thermal quantum discord of spins in an inhomogeneous magnetic field.
\newblock {\em Journal of Physics B: Atomic, Molecular and Optical Physics}, 44:065504, 2011.

\bibitem{10.1007/s10701-006-9065-9}
A.~Hagar and M.~Hemmo.
\newblock Explaining the unobserved—why quantum mechanics ain’t only about information.
\newblock {\em Foundations of Physics}, 36:1295--1324, 2006.

\bibitem{10.36347/sjpms.2020.v07i08.001}
Q.~Han, Z.~Lu, and Y.~Han.
\newblock Related properties of quantum measurements.
\newblock {\em Scholars Journal of Physics, Mathematics and Statistics}, 7:126--130, 2021.

\bibitem{10.32388/qj8nvr.2}
I.~S. Helland.
\newblock An alternative foundation of quantum theory.
\newblock 2023.

\bibitem{10.1111/0591-2385.00293}
C.~S. Helrich.
\newblock Measurement and indeterminacy in the quantum mechanics of dirac.
\newblock {\em Zygon®}, 35:489--503, 2000.

\bibitem{10.22533/at.ed.3174152421055}
A.~R. Hernández, A.~L.~E. Maldonado, and J.~H.~D. Bermejo.
\newblock Architecture of quantum computing.
\newblock {\em Journal of Engineering Research}, 4:2--10, 2024.

\bibitem{10.3389/fphy.2020.00139}
S.~Hossenfelder and T.~N. Palmer.
\newblock Rethinking superdeterminism.
\newblock {\em Frontiers in Physics}, 8, 2020.

\bibitem{10.1038/nature13460}
M.~Howard, J.~Wallman, V.~Veitch, and J.~Emerson.
\newblock Contextuality supplies the ‘magic’ for quantum computation.
\newblock {\em Nature}, 510:351--355, 2014.

\bibitem{10.1088/0253-6102/56/1/14}
M.~Hua, X.~Xiao, and Y.~Gao.
\newblock Screening effect in charge qubit.
\newblock {\em Communications in Theoretical Physics}, 56:74--78, 2011.

\bibitem{10.1002/que2.45}
W.~Huang, W.~Chien, C.~Cho, C.~Huang, T.~Huang, and C.~Chang.
\newblock Mermin's inequalities of multiple qubits with orthogonal measurements onibmq 53‐qubit system.
\newblock {\em Quantum Engineering}, 2, 2020.

\bibitem{10.1088/0031-8949/79/06/065013}
A.~Ibort, V.~I. Man’ko, G.~Marmo, A.~Simoni, and F.~Ventriglia.
\newblock An introduction to the tomographic picture of quantum mechanics.
\newblock {\em Physica Scripta}, 79:065013, 2009.

\bibitem{10.3390/photonics9020058}
D.~Im and Y.~Kim.
\newblock Decoherence-induced sudden death of entanglement and bell nonlocality.
\newblock {\em Photonics}, 9:58, 2022.

\bibitem{10.4236/jqis.2011.12005}
S.~Ishikawa.
\newblock A new interpretation of quantum mechanics.
\newblock {\em Journal of Quantum Information Science}, 01:35--42, 2011.

\bibitem{10.1021/acsami.7b09805}
W.Jang, J.Lee, C.In, H.Choi, and A.Soon.
\newblock Designing two-dimensional dirac heterointerfaces of few-layer graphene and tetradymite-type sb2te3 for thermoelectric applications.
\newblock {\em ACS Applied Materials \&Amp; Interfaces}, 9:42050--42057, 2017.

\bibitem{10.1117/12.620302}
P.Jorrand and S.Perdrix.
\newblock Unifying quantum computation with projective measurements only and one-way quantum computation.
\newblock {\em SPIE Proceedings}, 2005.

\bibitem{10.48550/arxiv.2002.10133}
M.Kantner and T.Koprucki.
\newblock Non-isothermal scharfetter-gummel scheme for electro-thermal transport simulation in degenerate semiconductors.
\newblock 2020.

\bibitem{10.1038/nature07127}
H.J. Kimble.
\newblock The quantum internet.
\newblock {\em Nature}, 2008.

\bibitem{10.1098/rspa.1998.0166}
E.Knill, R.Laflamme, and W.H. Zurek.
\newblock Resilient quantum computation: error models and thresholds.
\newblock {\em Proceedings of the Royal Society of London. Series A: Mathematical, Physical and Engineering Sciences}, 454:365--384, 1998.

\bibitem{10.48550/arxiv.2103.05301}
A.R. Kuzmak.
\newblock Measuring distance between quantum states on a quantum computer.
\newblock 2021.

\bibitem{10.48550/arxiv.2003.01011}
A.~R. Kuzmak and V.~M. Tkachuk.
\newblock Detecting entanglement by the mean value of spin on a quantum computer.
\newblock 2020.

\bibitem{10.1103/physrevlett.101.200501}
B.P. Lanyon, M.Barbieri, M.P. Almeida, and A.~G. White.
\newblock Experimental quantum computing without entanglement.
\newblock {\em Physical Review Letters}, 101, 2008.

\bibitem{10.1142/10541}
Nick Laskin.
\newblock Fractional quantum mechanics.
\newblock 2017.

\bibitem{10.1103/physreva.95.022334}
K.Li, G.Long, H.Katiyar, T.Xin, G.Feng, D.Lu, and R.Laflamme.
\newblock Experimentally superposing two pure states with partial prior knowledge.
\newblock {\em Physical Review A}, 95, 2017.

\bibitem{10.1007/s10773-010-0603-z}
G.Long.
\newblock Duality quantum computing and duality quantum information processing.
\newblock {\em International Journal of Theoretical Physics}, 50:1305--1318, 2010.

\bibitem{10.1557/s43577-021-00133-0}
V.Lordi and J.M. Nichol.
\newblock Advances and opportunities in materials science for scalable quantum computing.
\newblock {\em MRS Bulletin}, 46:589--595, 2021.

\bibitem{10.1126/science.1501466}
D.H. Mahler, L.A. Rozema, K.Bonsma-Fisher, L.Vermeyden, K.J. Resch, H.M. Wiseman, and A.M. Steinberg.
\newblock Experimental nonlocal and surreal bohmian trajectories.
\newblock {\em Science Advances}, 2:e1501466--e1501466, 2016.

\bibitem{10.1088/0305-4470/35/3/315}
O.~V. Man’ko, V.~I. Man’ko, and G.~Marmo.
\newblock Alternative commutation relations, star products and tomography.
\newblock {\em Journal of Physics A: Mathematical and General}, 35:699--719, 2002.

\bibitem{10.1007/s00453-002-0971-8}
Michele Mosca and Alain Tapp.
\newblock Introduction.
\newblock {\em Algorithmica}, 2002.

\bibitem{10.4028/www.scientific.net/amr.267.757}
D.~X. Nan, Y.~Zhang, and X.~Q. Sun.
\newblock Modeling of proportional integral derivative neural networks based on quantum computation.
\newblock {\em Advanced Materials Research}, 267:757--761, 2011.

\bibitem{10.1038/nature09418}
M.~Neeley, R.~C. Bialczak, M.~Lenander, E.~Lucero, M.~Mariantoni, A.~D. O’Connell, D.~Sank, H.~Wang, M.~Weides, J.~Wenner, Y.~Yin, T.~Yamamoto, A.~N. Cleland, and J.~M. Martinis.
\newblock Generation of three-qubit entangled states using superconducting phase qubits.
\newblock {\em Nature}, 467:570--573, 2010.

\bibitem{10.1103/physreva.75.012337}
M.~V.~d. Nest, W.~Dür, G.~Vidal, and H.~J. Briegel.
\newblock Classical simulation versus universality in measurement-based quantum computation.
\newblock {\em Physical Review A}, 75, 2007.

\bibitem{10.1088/1367-2630/17/3/033037}
A.~Nicolas, L.~Veissier, E.~Giacobino, D.~Maxein, and J.~Laurat.
\newblock Quantum state tomography of orbital angular momentum photonic qubits via a projection-based technique.
\newblock {\em New Journal of Physics}, 17:033037, 2015.

\bibitem{10.1038/nphoton.2013.355}
A.~Nicolas, L.~Veissier, L.~Giner, E.~Giacobino, D.~Maxein, and J.~Laurat.
\newblock A quantum memory for orbital angular momentum photonic qubits.
\newblock {\em Nature Photonics}, 8:234--238, 2014.

\bibitem{10.1007/s10702-005-1128-1}
H.~Nikolić.
\newblock Relativistic quantum mechanics and the bohmian interpretation.
\newblock {\em Foundations of Physics Letters}, 18:549--561, 2005.

\bibitem{10.1007/s10702-006-1009-2}
H.~Nikolić.
\newblock Classical mechanics without determinism.
\newblock {\em Foundations of Physics Letters}, 19:553--566, 2006.

\bibitem{10.1103/physrevlett.116.110403}
M.~Oszmaniec, A.~Grudka, M.~Horodecki, and A.~Wójcik.
\newblock Creating a superposition of unknown quantum states.
\newblock {\em Physical Review Letters}, 116, 2016.

\bibitem{10.1088/1742-6596/845/1/012030}
P.~O’Hara.
\newblock A generalized spin statistics theorem.
\newblock {\em Journal of Physics: Conference Series}, 845:012030, 2017.

\bibitem{10.1126/science.1226719}
A.~Peruzzo, P.~Shadbolt, N.~Brunner, S.~Popescu, and J.~L. O’Brien.
\newblock A quantum delayed-choice experiment.
\newblock {\em Science}, 338:634--637, 2012.

\bibitem{10.48550/arxiv.1712.09289}
A.~Poremba.
\newblock Quantum learning algorithms and post-quantum cryptography.
\newblock 2017.

\bibitem{10.1142/s0219749909005699}
R.~Raussendorf.
\newblock Measurement-based quantum computation with cluster states.
\newblock {\em International Journal of Quantum Information}, 07:1053--1203, 2009.

\bibitem{10.48550/arxiv.1510.06649}
M.~Rennela.
\newblock Operator algebras in quantum computation.
\newblock 2015.

\bibitem{10.1109/aucc.2016.7868220}
S.~Roy and I.~R. Petersen.
\newblock Robust h coherent-classical estimation of linear quantum systems.
\newblock {\em 2016 Australian Control Conference (AuCC)}, 2016.

\bibitem{10.3389/fphy.2024.1456491}
S.~K. Sahu and K.~Mazumdar.
\newblock State-of-the-art analysis of quantum cryptography: applications and future prospects.
\newblock {\em Frontiers in Physics}, 12, 2024.

\bibitem{10.1007/s11801-008-7153-0}
X.~Shang, L.~Yu, X.~Feng, and B.~Yang.
\newblock Generation and detection of squeezed state.
\newblock {\em Optoelectronics Letters}, 4:231--233, 2008.

\bibitem{10.1088/1751-8121/ac3bea}
F.~Shi, M.~Li, L.~Chen, and X.~Zhang.
\newblock Strong quantum nonlocality for unextendible product bases in heterogeneous systems.
\newblock {\em Journal of Physics A: Mathematical and Theoretical}, 55:015305, 2021.

\bibitem{10.1088/0143-0807/37/5/055706}
Pascal Sleutel, Erik Dietrich, Jan~van der Veen, and Wouter~van Joolingen.
\newblock Bouncing droplets: A classroom experiment to visualize wave-particle duality on the macroscopic level.
\newblock {\em European Journal of Physics}, 37:055706, 2016.

\bibitem{10.1038/ncomms3828}
X.~Su, S.~Hao, X.~Deng, L.~Ma, M.~Wang, X.~Jia, C.~Xie, and K.~Peng.
\newblock Gate sequence for continuous variable one-way quantum computation.
\newblock {\em Nature Communications}, 4, 2013.

\bibitem{10.1088/0305-4470/35/25/305}
V.~E. Tarasov.
\newblock Quantum computer with mixed states and four-valued logic.
\newblock {\em Journal of Physics A: Mathematical and General}, 35:5207--5235, 2002.

\bibitem{10.1103/physrevlett.119.230401}
T.~Theurer, N.~Killoran, D.~Egloff, and M.~B. Plenio.
\newblock Resource theory of superposition.
\newblock {\em Physical Review Letters}, 119, 2017.

\bibitem{10.1364/ol.43.005765}
C.~Yang, Z.~Zheng, and Y.~Zhang.
\newblock Universal quantum gate with hybrid qubits in circuit quantum electrodynamics.
\newblock {\em Optics Letters}, 43:5765, 2018.

\bibitem{10.21203/rs.3.rs-475920/v1}
Chen Yang and S.~Olutunde Oyadiji.
\newblock Modelling wave-particle duality of classical particles and waves.
\newblock 2021.

\bibitem{10.1088/1402-4896/ada313}
Xiaowan Yang.
\newblock Experimental verification of a tighter wave-particle duality based on quantum coherence.
\newblock {\em Physica Scripta}, 100:025110, 2025.

\bibitem{10.1364/fio.2017.fth3e.6}
S.~Yokoyama, N.~D. Pozza, T.~Serikawa, K.~B. Kuntz, T.~A. Wheatley, D.~Dong, E.~H. Huntington, and H.~Yonezawa.
\newblock The quantum entanglement of measurement.
\newblock {\em Frontiers in Optics 2017}, page FTh3E.6, 2017.

\bibitem{10.1016/j.physleta.2008.09.053}
V.~I. Yukalov and D.~Sornette.
\newblock Quantum decision theory as quantum theory of measurement.
\newblock {\em Physics Letters A}, 372:6867--6871, 2008.

\bibitem{10.48550/arxiv.0904.4021}
Wenzhuo Zhang.
\newblock Particle-like property of vacuum states.
\newblock 2009.

\bibitem{10.32474/mams.2020.03.000157}
B.~Zohuri and F.~Rahmani.
\newblock What is quantum computing and how it works, artificial intelligence driven by quantum computing.
\newblock {\em Modern Approaches on Material Science}, 3, 2020.

\end{thebibliography}

\end{document}